\documentclass[journal]{IEEEtran}
\ifCLASSINFOpdf
\else
\fi
\usepackage{graphicx}
\graphicspath{{/Figures/}}
\DeclareGraphicsExtensions{.eps,.pdf,.png,.jpg}
\usepackage{subfig}
\usepackage{amsmath}
\usepackage{algorithm}
\usepackage{algorithmic}
\usepackage{cite}
\usepackage{amsthm}
\usepackage{amssymb}
\usepackage{color}
\usepackage{multicol}
\usepackage{bm}
\usepackage{tabularx}
\usepackage{booktabs}
\usepackage{balance}
\usepackage[top=0.75in, bottom=0.85in, left=0.635in, right=0.635in]{geometry}

\theoremstyle{definition}

\theoremstyle{remark}

\theoremstyle{proposition}

\begin{document}
\title{Machine-to-Machine (M2M) Communications in Virtualized Cellular Networks with MEC}

\author{\IEEEauthorblockN{Meng Li\IEEEauthorrefmark{1} and S. Hu\IEEEauthorrefmark{2}}\\
\thanks{This work is supported by the National Natural Science Foundation of China under Grants No. 61372089, No. 61571021  and No. 61671029.}
\IEEEauthorblockA{\IEEEauthorrefmark{1}Faculty of Information Tech., Beijing Univ. of Tech., Beijing, P.R.~China}\\
\IEEEauthorblockA{\IEEEauthorrefmark{2}Depart. of Systems and Computer Eng., Carleton Univ., Ottawa, ON, Canada}\\
}
\maketitle

% As a general rule, do not put math, special symbols or citations in the abstract or keywords.
\begin{abstract}
As an important part of the Internet-of-Things (IoT), machine-to-machine (M2M) communications have attracted great attention. In this paper, we introduce mobile edge computing (MEC) into virtualized cellular networks with M2M communications, to decrease the energy consumption and optimize the computing resource allocation as well as improve computing capability. Moreover, based on different functions and quality of service (QoS) requirements, the physical network can be virtualized into several virtual networks, and then each MTCD selects the corresponding virtual network to access. Meanwhile, the random access process of MTCDs is formulated as a partially observable Markov decision process (POMDP) to minimize the system cost, which consists of both the energy consumption and execution time of computing tasks. Furthermore, to facilitate the network architecture integration, software-defined networking (SDN) is introduced to deal with the diverse protocols and standards in the networks. Extensive simulation results with different system parameters reveal that the proposed scheme could significantly improve the system performance compared to the existing schemes.\
\end{abstract}

% Note that keywords are not normally used for peer review papers.

% \fi
\IEEEpeerreviewmaketitle

\section{Introduction}\label{sec:Introduction}
Machine-to-machine (M2M) communications, also named as machine-type communications (MTCs), have attracted great attention both academia and industry~\cite{WC15,IT14}. Unlike traditional wireless communications, M2M devices are typically equipped with limited resources for a relatively long working life~\cite{IT14,GC15}. Therefore, in many M2M applications, the energy consumption saving gets more and more imperative than the throughput increasement, since more MTCDs tend to transmit small data with limited energy~\cite{GC15,HH12}.

Many research efforts have been conducted to improve the performance of energy consumption in M2M communications and in wireless communication in general \cite{XYJL12,BYC12}. The authors of~\cite{PK15} investigated a novel medium access control (MAC) protocol with low latency and energy efficiency in hierarchical M2M networks, in order to accommodate efficient data transmission from a terminal node to a sink node via cluster heads. The authors of~\cite{LW15} presented a novel scheme in M2M-based home environment, and the energy savings are formulated into an optimization problem to minimize the total energy consumption, even under multiple user comfort constraints. An energy-efficient data aggregation scheme for a hierarchical M2M network was proposed in~\cite{MD16}, and the authors developed a coverage probability-based optimal data aggregation scheme for M2M devices to minimize the average total energy expenditure.\

Another important issue in M2M communications is computation. Many resource-constrained MTCDs are not be able to rely solely on their own limited resources to fulfill their computing needs~\cite{CZ16}. Traditionally, to address the computational capability issue, mobile cloud computing (MCC) systems have been extensively studied~\cite{CYB14,YYB15}. Nevertheless, as the distance between the cloud and the MTCD is usually large, MCC may not provide guarantees to low latency applications (e.g., emergency services), and frequent transmitting data (e.g., location information) from the MTCD to the cloud may not be feasible or economical~\cite{AFG15}. Moreover, requiring all of MTCDs to interact directly with the cloud will be unrealistic and cost prohibitive since it often requires resource-intensive processing and complex protocols~\cite{CZ16}. In order to tackle these issues, a novel technique, called mobile edge computing (MEC), is being standardized to allocate computing resources in wireless cellular networks~\cite{ETSI14}. MEC allows MTCDs to perform computation offloading to offload their computing tasks to the MEC server via wireless cellular networks.\

Although some excellent works have been done on the energy consumption and computation in M2M communications, these two important aspects were generally considered separately in the existing works. In this paper, we propose a novel framework to jointly consider both energy consumption and computation in M2M communications. Firstly, MEC is introduced into M2M communication networks, and the computing tasks of MTCDs can be offloaded to the MEC server, then the network can accommodate more MTCDs with low energy consumption. Moreover, wireless network virtualization (WNV) and software-defined networking (SDN) are applied in the proposed framework. WNV enables a physical wireless network to abstract and slice into multiple virtual ones~\cite{LY15}, such that differentiated M2M services can be offered differentiated QoS requirements. Meanwhile, the SDN paradigm is introduced to integrate diverse protocols and standards of MEC, WNV and M2M communications. In addition, we formulate the random access in M2M communications as a partially observable Markov decision process (POMDP). Simulation results are presented to show the performance improvement of the proposed scheme.

The rest of this paper is organized as follows. The system model is presented in Section \ref{sec:Systemmodel}. In Section \ref{sec:Algorithm1}, the random access is formulated. Section \ref{sec:Simulation} discusses the simulation results. Finally, we conclude this work in Section \ref{sec:Conclusion}.\

\section{System Model}\label{sec:Systemmodel}
In this section, we introduce the network model for M2M communications. Then, the computing model is also described, followed by the description of energy consumption model.\

\subsection{Network Model}
\begin{figure}[!t]
\centering
\includegraphics[width=3.6in]{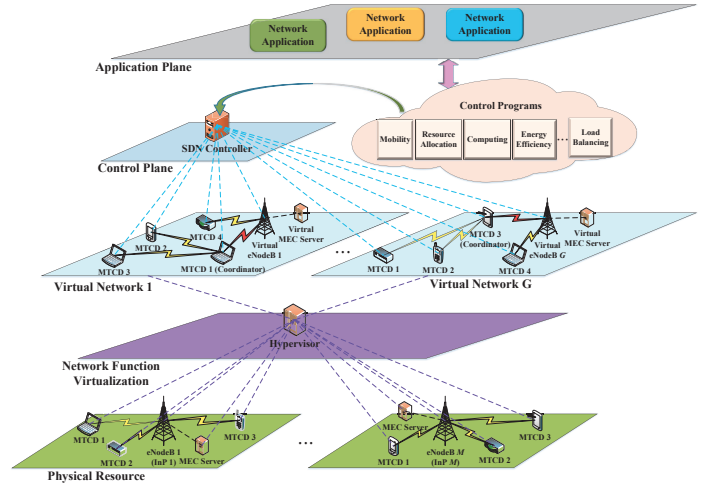}
 %where an .eps filename suffix will be assumed under latex,
% and a .pdf suffix will be assumed for pdflatex; or what has been declared
% via \DeclareGraphicsExtensions.
\caption{An architecture of virtualized and software-defined cellular networks with M2M communications.}
\label{fig:model}
\end{figure}

An example of the virtualized and software-defined cellular networks with M2M communications is depicted in Fig. \ref{fig:model}, the preferred network architecture is given in detail as follows.

\subsubsection{Physical Resource Layer}
The physical resource layer, which includes eNodeB, resource blocks (RBs), MEC servers, power from different infrastructure providers (InPs), etc., is responsible for providing available physical resources. In the proposed framework, there are totally $M$ InPs offering wireless access services. Meanwhile, the total number of MTCDs is $N$, and the $m$-th InP deploys the $m$-th cellular network, which possesses $N_m$ ($1\leq N_m\leq N$) MTCDs and one eNodeB~\cite{CY16}.

\subsubsection{Wireless Network Virtualization}
The hypervisor is an important component in WNV. In general, the hypervisor is typically deployed in the physical eNodeB, and could provide functions to connect physical resources with virtual eNodeBs~\cite{LY15}. Through WNV, the physical network would be virtualized into several virtual networks with M2M communications, based on different functions or QoS requirements. Handover \cite{MYL04,YK07} and node mobility \cite{YL01} are not considered due to simplicity.\

\subsubsection{Virtual Network Layer}
In this layer, the $g$-th virtual network consists of one virtual eNodeB with virtual MEC server, and $N_g$ ($1\leq N_g \leq N$) MTCDs~\cite{LY16}. In the proposed framework, one MTCD will be assigned as the coordinator in the group, denoted as $n_{coor}$. Meanwhile, other MTCDs are denoted as $n_1, n_2,\dots, n_x\dots, n_{N_{g}-1}$, and all of them belong to the set $\mathcal{N}_{g,ue}$. Similar to spectrum sensing in cognitive radios \cite{LYH10}, each MTCD will sense the RB, decide to access the eNodeB or coordinator at the beginning of each time slot, and eventually make decisions on the transmission behaviour.\

As in~\cite{TH11}, the coordinator $n_{coor}$ in the $g$-th virtual network can be determined by an existing scheme named as ``A-means''. As such, the coordinator, which has the maximum arithmetic mean off channel gain to other MTCDs in the virtual network, can be calculated as

\begin{equation}
n_{coor}=\arg \max_{n_{x}}\left\{\frac{1}{N_{g}-1}\sum\limits_{n_{x}\neq n_{y}} h_{n_x, n_y}\right\}, \forall {n_y},
\end{equation}
where $h_{n_x, n_y}$ denotes the channel gain from the $n_x$-th MTCD to the $n_y$-th MTCD.\

The MTCDs access problem is considered with equal-sized time frames, and each time frame is further divided into $K$ time slots. Let the total number of RBs offered by the physical eNodeB be $R_{total}$. Through network virtualization, for the $g$-th virtual network, the number of RBs for MTCDs is $R_g$ ($1 \leq R_g \leq R_{total}$). At each initial time slot $\delta t_{k}$, the MTCD attempts to access to either eNodeB directly, or the coordinator $n_{coor}$ in the group. The number of RBs to access the eNodeB or the coordinator is represented as $R_{g,1}$ and $R_{g,2}$, respectively. Obviously, $R_{g, 1}+R_{g, 2}=R_{g}$ holds.\

Let $\bm{s_r}=\{0,1\}$ be the state of the $r$-th RB, where $0$ represents idle while $1$ just the opposite. In the uplink, for the $n_x$-th MTCD accessing to the $r$-th RB, the transmission rate available in time slot $\delta t_{k}$ can be calculated as
 \begin{equation}
C_{n_x,r}(k)=\left\{
\begin{array}{lcl}
B_{n_x,r} \log_2 \left\{ 1+\frac{P_{n_x}(k)h_{n_x,r}}{\sigma^{2}}\right\},\\ \qquad\qquad\qquad\qquad\qquad\quad \quad\ \text{if}\ s_{r}=0,\\
B_{n_x,r} \log_2 \left\{1+\frac{P_{n_x}(k)h_{n_x,r}}{\sum\limits_{n_{y}\neq n_{x},n_{y}\in \mathcal{N}_{g,ue}} P_{n_y}(k)h_{n_{y},r}+\sigma^{2}}\right\},\\ \qquad\qquad\qquad\qquad\qquad\quad \quad\ \text{if}\ s_{r}=1,&\\
\end{array}
\right.
\end{equation}
where $B_{n_x,r}$ represents the bandwidth offered by the $r$-th RB, $P_{n_x}(k)$ ($P_{n_y}(k)$) is the transmit power consumed on the $r$-th RB by the $n_x$-th ($n_y$-th) MTCD, $h_{n_x,r}$ ($h_{n_{y},r}$) is the channel gain of the $n_x$-th ($n_{y}$-th) MTCD on the $r$-th RB, and $\sigma^{2}$ is the system noise power.\

In addition, the SDN controller will allocate additional RBs for the link between the coordinator and the virtual eNodeB, the number of these RBs could be denoted as $R_{g}^{'}$. The transmission rate between the the coordinator $n_{coor}$ and virtual eNodeB can be calculated as

\begin{equation}
C_{n_{coor},r^{'}}(k)=B_{n_{coor},r^{'}} \log_2 \left\{1+\frac{P_{n_{coor}}(k)h_{n_{coor},r^{'}}}{\sigma^{'2}}\right\},
\end{equation}
where $B_{n_{coor},r^{'}}$ is the bandwidth provided by the coordinator $n_{coor}$, $P_{n_{coor}}(k)$ is the transmit power consumed on the $r^{'}$-th RB by the coordinator, $h_{n_{coor},r^{'}}$ is the channel gain of the coordinator on the $r^{'}$-th RB, and $\sigma^{'2}$ is the system noise power.\

\subsubsection{Controller Layer}
According to different QoS requirements, the SDN controller would allocate and adjust RBs in M2M communication networks. Besides, diverse protocols and standards can be managed and integrated by the SDN controller.\

 \subsubsection{Application Layer}
It consists of multiple network applications. The goal of network design is to implement and fulfill these network applications.\

\subsection{Computing Model}\
In each virtual network, the capabilities to handle computing tasks exist in MTCDs, the coordinator and even MEC servers. In the proposed framework, suppose that the $n_x$-th MTCD has to execute computing task $\mathcal{I}_{n_{x}}\triangleq (\alpha_{n_{x}}, \beta_{n_{x}})$, where $\alpha_{n_{x}}$ is the size of input data involved and $\beta_{n_{x}}$ represents the total number of CPU cycles required to accomplish computing tasks~\cite{C15}. In what follows, the model of both local computing and MEC will be discussed respectively.\

\subsubsection{Local Computing}
The computing capability (i.e., CPU cycles per second) of the $n_x$-th MTCD could be represented as $F_{n_x}^l$, and it can be reflected by the execution time, i.e.,
\begin{equation}
t_{n_x}^l(k)=\frac{\beta_{n_x}(k)}{F_{n_x}^l}.
\end{equation}

However, if the $n_x$-th MTCD selects to access the coordinator $n_{coor}$, the computing task will be executed on the coordinator, the computing capability of which can be represented as $F_{n_{coor}}^{l^{'}}$. In this case, the computing task has to be offloaded at first, and then be transferred to the coordinator through wireless communication link. The transmission time can be denoted by
\begin{equation}
t_{n_x,off}^{l'}(k)=\frac{\alpha_{n_x}(k)}{C_{n_{x},r}(k)}.
\end{equation}

After offloading, the coordinator will execute the computing task and the execution time is represented as
\begin{equation}
t_{n_x,comp}^{l^{'}}(k)=\frac{\beta_{n_x}(k)}{F_{n_{coor}}^{l^{'}}}.
\end{equation}

Therefore, in the case that the coordinator handles the computing task, the total execution time turns out to be
\begin{equation}
t_{n_x}^{l^{'}}(k)=t_{n_x,off}^{l'}(k)+t_{n_x,comp}^{l^{'}}(k).
\end{equation}

\subsubsection{Mobile Edge Computing}
For the MEC, the computing task needs to be offloaded to the MEC server. In particular, the MEC server will divide the whole computing task $\mathcal{I}_{n_x}$ into two steps: the computing task offloading and the input data transfer to the MEC server. As a result, the transmission time of the $n_x$-th MTCD to offload the input data can be calculated by
\begin{equation}
t_{n_x,off}^c(k)=\frac{\alpha_{n_x}(k)}{C_{n_x,r}(k)}.
\end{equation}

After data offloading, the MEC server will execute the computing task $\mathcal{I}_{n_x}$. The computing capability of the MEC server can be denoted by $F_{mec}^c$, and the time to execute the computing task on the MEC server turns out to be
\begin{equation}
t_{n_x,comp}^c(k)=\frac{\beta_{n_x }(k)}{F_{mec}^c}.
\end{equation}

Therefore, the overhead of the MEC approach in terms of the processing time can be represented as
\begin{equation}
t_{n_x}^c(k)=t_{n_x,off}^c(k)+t_{n_x,comp}^c(k).
\end{equation}

Finally, in time slot $\delta t_k$, the total execution time of the computing task can be represented as

\begin{equation}\label{time}
t_{{n_x},total}(k)=\left\{
\begin{array}{ll}
t_{n_x}^l(k),& \text{for the local device},\\
t_{n_x}^{l^{'}}(k),& \text{for the coordinator},\\
t_{n_x}^c(k),& \text{for the MEC server}.\\
\end{array}
\right.
\end{equation}

\subsection{Energy Consumption Model}\
In M2M communications, energy efficiency has been considered as an important factor from two aspects: one is the energy consumption in the access phase, and the other one is that in the computation task. For the energy consumption in the access process, it could be divided into two parts: RBs access sensing and information transmission. The energy consumptions for the RB sensing and information transmission by the $n_x$-th MTCD are represented as $P_{n_{x}}^{'}$ and $P_{n_x}$, respectively. In addition, let the time for RB sensing and data transmission be $t_{se}$ and $t_{tr}$, respectively. We assume that it only has one packet to send, and the fixed packet size is $D_{n_x}$. Therefore, the transmission time $t_{tr}(k)$ can be calculated by
\begin{equation}
t_{tr}(k)=\left\{
\begin{array}{lcl}
\frac{D_{n_x}(k)}{C_{n_x,r}(k)}, &\text{if the MTCD access the eNodeB},& \\ \\
\frac{D_{n_x}(k)}{C_{n_{coor},r^{'}}(k)}, &\text{if the MTCD access the coordinator}.&
\end{array}
\right.
\end{equation}

As a result, the energy consumption used for sensing and data transmission in time slot $\delta t_k$ is given as
\begin{equation}
E_{st}(k)=\left\{
\begin{array}{lcl}
0,\ \ \ \ \ \ \ \ \ \ \ \ \ \ \ \ \ \ \ \ \ \ \ \ \ \text{if there exists no sensing},\\
P_{n_x}^{'}\cdot t_{se}(k),\ \ \ \ \ \ \ \ \ \ \ \ \ \ \ \ \ \ \text{if one RB is sensed},\\
P_{n_x}\cdot t_{tr}(k)+P_{n_{x}}^{'}\cdot t_{se}(k),\ \text{if one RB is accessed}.\\
\end{array}
\right.
\end{equation}

For the energy consumption of data computation tasks, if one MTCD decides to execute the computation task on the local device, the energy consumption can be represented as
\begin{equation}
E_{n_{x}}^l(k)=e_{n_x}\beta_{n_x}(k),
\end{equation}
where $e_{n_x}$ is the coefficient denoting the consumed energy per CPU cycle on the $n_x$-th MTCD. As in~\cite{C15} and~\cite{WZ12}, this coefficient can be set as $e_{n_{x}}\approx 10^{-11}(F_{n_{x}}^l)^2$.\

Focusing on the computation task executed on the coordinator $n_{coor}$, since the input data needs to be offloaded and transmitted by the $n_x$-th MTCD, the energy consumption can be calculated by
\begin{equation}
E_{n_x}^{l^{'}}(k)=\frac{P_{n_x}\alpha_{n_x}(k)}{C_{coor,r^{'}}(k)}.
\end{equation}

In addition, for the computation task operated on the MEC server, the energy consumption used for offloading and transmitting the input data by the $n_x$-th MTCD turns out to be
\begin{equation}
E_{n_x}^c(k)=\frac{P_{n_x}\alpha_{n_x}(k)}{C_{{n_x},r}(k)}.
\end{equation}

As such, the total energy consumption is represented as

\begin{equation}\label{energy}
E_{{n_x},total}(k)=\left\{
\begin{array}{ll}
E_{st}(k)+E_{n_x}^l(k),& \text{select local device},\\
E_{st}(k)+E_{n_x}^{l^{'}}(k),& \text{select the coordinator},\\
E_{st}(k)+E_{n_x}^c(k),& \text{select MEC server}.\\
\end{array}
\right.
\end{equation}

\section{A Solution to Random Access, Energy Consumption and Computation Node Selection in M2M Communications}\label{sec:Algorithm1}
In this section, we present a new stochastic optimization method for solving the problem of random access with M2M communications. Then, each tuple of POMDP is described and given in detail, followed by the algorithm of system reward and optimization object, respectively.\

\subsection{POMDP Formulation}
Since the state of RBs cannot be directly and accurately observed by MTCDs, the random access optimization problem with the minimum system costs can be easily formulated as a POMDP~\cite{LM14}. For simplicity, the POMDP formulation is discussed by taking the $g$-th virtual network as an example.\

\emph{1) Action Space} \

The action space is considered as a combined space, where the RB sensing selection and decision, the access node selection as well as the computing node selection coexist. In time slot $\delta t_k$, the MTCD accessing the network has to execute these actions: \emph{Sensing Decision}, \emph{Access Decision} and \emph{Computing Node Selection}. Thus, the composite action $a(k)\in\mathcal{A}$ can be denoted as

\begin{equation}
a(k)=\{a_s(k), a_a(k), a_c(k)\},
\end{equation}
where $a_s(k)$ denotes the RBs sensing action, $a_a(k)$ denotes the access decision, and $a_c(k)$ denotes the selection of computing node, respectively.\

\textbf{Sensing Decision:} Let $a_s(k)$ be the set of RB sensing decisions and actions in each time slot, it can be defined as
\begin{equation}
\begin{array}{ll}
a_{s}(k) \in \{0 (no\ sensing), {RB_1},\dots, {RB_r},\dots, {RB_{R_g}}\}.
\end{array}
\end{equation}

For $a_{s}(k)$, $0$ represents that the MTCD will not sense RBs and select sleep mode, $RB_r$ represents that the MTCD will select the $r$-th RB to sense.\

\textbf{Access Decision:} After the RB sensing, the MTCD will decide whether or not to access the network. Due to the existing coordinator in the virtual network, the $n_x$-th MTCD has two choices: either the eNodeB directly or the coordinator. The access decision $a_a(k)$ could be defined as
\begin{equation}
a_{a}(k) \in \{{0}, {1}, {2}\},
\end{equation}
where $0, 1, 2$ represents that the MTCD do not access the RB, accesses the eNodeB and accesses the coordinator, respectively.\

\textbf{Computing Node Selection:} When one RB is selected by the $n_x$-th MTCD, the corresponding computing node will be determined based on the access decision. The decision of computing node selection $a_c(k)$ can be represented as
\begin{equation}
a_{c}(k) \in \{{0}, {1}, {2}\},
\end{equation}
where $0$ represents that the computing task will be executed on the MTCD, $1$ represents that the MEC server will be selected to handle the computing task, and $2$ implies that the coordinator will execute the computing task offloaded by the MTCD.\

\emph{2) State Space and Transition Probability}\

The system state space $\mathcal{S}$ is the set of all RB states, and the state in time slot $\delta t_k$ can be denoted as
\begin{equation}
s(k)=[{s_{1}(k)}{s_{2}(k)}\dots{s_{r}(k)}\dots{s_{R_g}(k)}],
 \end{equation}
 where $s(k)\in\mathcal{S}$. For the state of each RB, let $s_r(k)$ represent the $r$-th RB state, it can be defined as
\begin{equation}
\begin{aligned}
{s_{r}(k)} \in \{0 (idle),1(busy)\}.
\end{aligned}
\end{equation}

The one-step transition probability of the $r$-th RB state from time slot $\delta t_{k}$ to $\delta t_{k+1}$ is given by
 \begin{equation}\label{stateprobability}
\begin{aligned}
p_{ i,j}=Pr\{s_r(k+1)=j\mid s_r(k)=i\}, \forall i, j\in s_r,
\end{aligned}
\end{equation}
where $p_{i,j}$ is the transition probability of the RB state from state $i$ to state $j$.

\emph{3) Observation Space}\

Since it is intractable to obtain the full knowledge of each RB state directly, the MTCD needs to observe the RB state based on the state transition and optimal action taken in this time slot~\cite{LM14}. Let $\theta(k)\in\Theta$ be the composite observation state in time slot $\delta t_k$. Then $\theta(k)$ can be identified as
\begin{equation}
\begin{aligned}
{\theta(k)}=[{\widehat{s_{1}}(k)}{\widehat{s_{2}}(k)}\dots{\widehat{s_{r}}(k)}\dots{\widehat{s_{R_g}}(k)}].
\end{aligned}
\end{equation}

Focusing on the $r$-th RB, $\widehat{s_{r}}(k)$ is the observation state of $s_{r}(k)$, and can be written as
\begin{equation}
\begin{aligned}
{\widehat{s_{r}}(k)} \in \{0 (idle),1(busy)\}.
\end{aligned}
\end{equation}

Then, the probability of observation state is defined as $b^{a}_{s_{r},\widehat{s_{r}}}$, where $b^{a}_{s_{r},\widehat{s_{r}}}(k)=Pr\{\widehat{s_{r}}(k) \mid s_{r}(k), a(k)\}$. It is known when the RB state is $s_{r}(k)$ and composite action is $a(k)$ in the time slot $\delta t_k$, therefore, it can be calculated as
\begin{eqnarray}
%\begin{displaymath}
b^{a}_{s_{r},\widehat{s_{r}}}(k)=\left\{
\begin{array}{lll}
\nu, &\text {if}~a(k)=RB_r,~\widehat{s_{r}}(k)=0,\\
1-\nu, &\text {if}~a(k)=RB_r,~\widehat{s_{r}}(k)=1,\\
\omega, &\text {if}~a(k)=0,~\widehat{s_{r}}(k)=0,\\
1-\omega, &\text {if}~a(k)=0,~\widehat{s_{r}}(k)=1,\\
\end{array}
\right.
\end{eqnarray}
%\end{displaymath}
where $\nu$ and $\omega$ are the probabilities of false detection.

\emph{4) Information State}\

Information state is considered as an important element in POMDP. Since all RB states may not be known directly by the MTCD, it can be obtained depend on its action decision and observation history encapsulated by the information state~\cite{LM14}. Let $\pi(k)=\{\pi_1^{k}, \pi_2^{k},\dots,\pi_{i}^{k},\dots,\pi_{j}^{k},\dots,\pi_{\mathcal{S}}^{k}\}$ (${i, j\in s_r}$) denote the information space, where $\pi_{i}^{k}\in[0,1]$ represents the conditional probability (given decision and observation history) that the current state is $i$ at the beginning of time slot $\delta t_k$. At the end of each time slot, the information state is updated through Bayes' rule~\cite{LM14}, and it can be represented as
\begin{equation}
\pi_{j}^{k+1}=\frac{\sum_{i}\pi_{i}^{k}p_{i, j}b^{a}_{j,\widehat{s_{r}}}(k)}{\sum_{i, j} \pi_{i}^{k}p_{i, j}b^{a}_{j,\widehat{s_{r}}}(k)}.
\end{equation}

\emph{5) Reward and Objective}\

In this paper, by regarding both the energy consumption and computing processing time as system rewards, the reduction of energy consumed by MTCDs and the execution time of computing tasks can be realized for performance evaluation. According to Eqs.~(\ref{time}) and (\ref{energy}), the functions related to the computing time and energy consumption can be represented as
\begin{equation}
Re_{n_x}(k)=\zeta t_{{n_x},total}(k)+\eta E_{{n_x},total}(k),
\end{equation}
where $\zeta$ and $\eta$ ($0\leq\zeta, \eta\leq1, \zeta+\eta=1$) represent the weight factors of the execution time and energy consumption, respectively. Then the expected total rewards can be defined as
\begin{equation}
Re=E_{\{\mu_{s}, \mu_{a}, \mu_{c}\}}\left[\sum\limits^{K-1}_{k=0}\sum\limits_{n_{x}\in\mathcal{N}_{g,ue}}Re_{n_x}(k)\right],
\end{equation}
where $\mu_{s}$ is the RB sensing policy that specifies the sensing decision $a_{s}$, $\mu_{a}$ is the RB access policy that specifies the access decision $a_{a}$, and $\mu_{c}$ is the computing node selection policy that specifies the node selection decision $a_{c}$. Moreover, $E_{\{\mu_{s}, \mu_{a}, \mu_{c}\}}$ indicates the expectation given that the policies $\mu_{s}$, $\mu_{a}$ and $\mu_{c}$ are employed. We aim to develop a joint design with an optimal policy set $\mathcal{U^{*}}$ for the system performance improvement. Hence, $\{\mu_{s}^{*}, \mu_{a}^{*}, \mu_{c}^{*}\}$ should be a joint policy that could maximize the expected total rewards in the decision time frame, i.e.,
\begin{equation}\label{reward}
\{{\mu_{s}^{*}}, {\mu_{a}^{*}}, {\mu_{c}^{*}}\}=\arg \min \limits_{\{\mu_{s}^{*}, \mu_{a}^{*}, \mu_{c}^{*}\}\in \mathcal{U^{*}}}E_{\{\mu_{s}, \mu_{a}, \mu_{c}\}}\left[Re\right].
\end{equation}

\subsection{The Solution to the POMDP Problem}
In this subsection, a dynamic programming is proposed to solve the POMDP problem, where a value function, defined over the entire information space, is introduced to derive the optimal policy~\cite{WYS10}. As the value function, $W_{k}(\pi({k}))$ stands for the minimum expected system cost that can be obtained from time slot $\delta t_k$, given information state $\pi(k)$. Assuming that the MTCD attempts to access the RB with action $a(k)$ and observation acknowledgement $\widehat{s_{r}}(k)$, the reward can be accumulated from time slot $\delta t_{k}$. As a result, the optimal random access policy can be calculated as
\begin{eqnarray}\label{totalreward}
\begin{split}
%\begin{array}
W_{k}(\pi(k))= \min\limits_{a(k)\in\mathcal{A}}\{\sum\limits_{i\in \mathcal{S}}\sum\limits_{j\in \mathcal{S}}\pi_{i}^{k}p_{i, j} \sum\limits_{\widehat{s_{r}}(k)\in \mathcal{S}}b^{a(k)}_{j,\widehat{s_{r}}(k)}\cdot\\ [Re_{n_x}(k)+W_{k+1}(\pi(k+1))]\} , \forall 1\leq k \leq K-1.
%\end{array}
\end{split}
\end{eqnarray}

%Obviously, in the real-world M2M communication networks, to derive the optimal policy $\{{\mu_{s}^{*}}, {\mu_{a}^{*}}, {\mu_{c}^{*}}\}$ at each time slot, it has to execute several complex computations. Fortunately, due to the introduction of the SDN controller, the calculation of the optimal policy can be performed off-line. Given the values of all parameters, the action decision process can be determined and stored in the table. The MTCDs will select the optimal action with corresponding current state.

\section{Simulation Results and Discussions}\label{sec:Simulation}
In this section, simulation results are presented to show the performance of the proposed scheme. We consider $M=3$ InPs offering wireless access services with M2M communications in a radius of 1 KM region, also including $3$ eNodeBs and $N=50$ randomly distributed MTCDs. After virtualization, all the physical networks can be sliced into $G=5$ virtual networks. Each virtual network consists of one virtual eNodeB and several MTCDs, one of which is selected as the coordinator according to channel conditions. In the initial time slot, the virtual eNodeB is allocated $5$ RBs. For the wireless access link, the channel bandwidth between the MTCD and virtual eNodeB or between the coordinator and virtual eNodeB is set as $5$ MHz and $10$ MHz, respectively. The transmission power of each MTCD is $100$ mWatts, while the system background noise power is $1$ mWatts. In addition, for the computing task, the fixed packet size is $D_{n_{x}}=2$ MB, the data size for the computation offloading is $\alpha_{n_{x}}=420$ KB, and the total number of CPU cycles is $\beta_{n_{x}}=1000$ Megacycles. The CPU computation capability of the MTCD, the coordinator and the MEC servers are set to be $F_{n_x}^l=0.5$ GHz, $F_{n_{coor}}^{l^{'}}=1$ GHz and $F_{mec}^{c}=100$ GHz, respectively.\

For the sake of simplicity, we only focus on one virtual network in the simulation. The RB state transition matrix is constructed by the probability. And the probability that RB remains idle state, remains busy state, transits from busy to idle state and transits from idle to busy state is set as $0.8$, $0.15$, $0.85$, and $0.2$, respectively. The probability with false observation is set as $\nu=\omega=0.1$. Each time frame includes $100$ time slots.\

\begin{figure}[!t]
\centering
\includegraphics[width=3in]{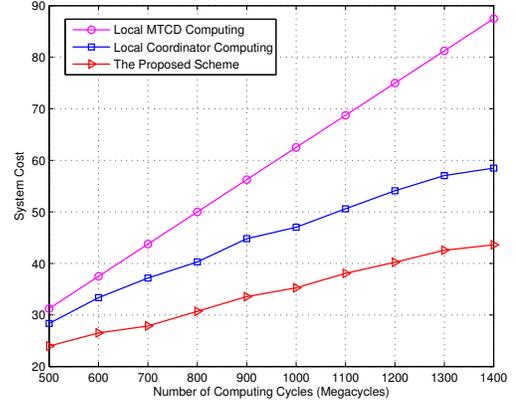}
 %where an .eps filename suffix will be assumed under latex,
% and a .pdf suffix will be assumed for pdflatex; or what has been declared
% via \DeclareGraphicsExtensions.
\caption{System cost with different numbers of computing cycles.}
\label{fig:computing1}
\end{figure}

Fig. \ref{fig:computing1} shows the system cost with different numbers of computing cycles. The system cost induced by local MTCD computing, local coordinator computing and the proposed scheme all increase with the growth of computing cycles. However, the system cost by the proposed scheme increases much slower than those by other two schemes. The advantage of the proposed scheme is prominent because MEC can be utilized by the MTCD through the proposed POMDP optimization strategy. Hence, with the increasing number of computing cycles, more MTCDs will select MEC to handle the computing tasks. Then the heavy cost of local computing can be mitigated and the system cost is decreased obviously.\

%\begin{figure}[!t]
%\centering
%\includegraphics[width=3in]{S2.eps}
% %where an .eps filename suffix will be assumed under latex,
%% and a .pdf suffix will be assumed for pdflatex; or what has been declared
%% via \DeclareGraphicsExtensions.
%\caption{System cost with different data sizes for computation offloading.}
%\label{fig:computing2}
%\end{figure}
%
%Fig. \ref{fig:computing2} depicts the variation of system cost with different data sizes for computation offloading. It can be seen that with the increasing number of data sizes, the system cost by both the proposed scheme and local coordinator computing grow obviously. However, the system cost in the proposed scheme is still lower than other two schemes. Especially, the smaller the data size for computation offloading, the more significant the advantage of proposed scheme. Moreover, the system cost by the proposed scheme increases slowly when the data size for computation offloading gets larger and larger. The reason is that the MTCD tends to select local computing or MEC via POMDP optimization strategy when the data size for computation offloading is large, in order to avoid the heavy cost of computation offloading through wireless transmission link.\

\begin{figure}[!t]
\centering
\includegraphics[width=3in]{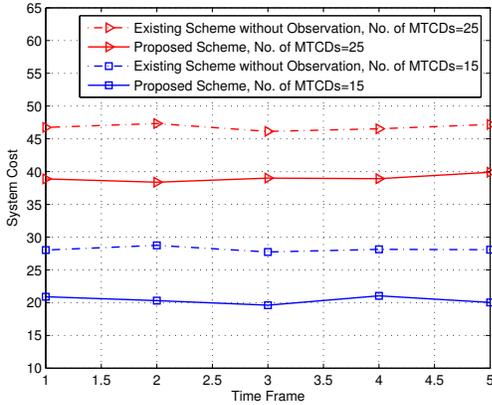}
 %where an .eps filename suffix will be assumed under latex,
% and a .pdf suffix will be assumed for pdflatex; or what has been declared
% via \DeclareGraphicsExtensions.
\caption{System cost with different numbers of MTCDs in different time frames.}
\label{fig:convergence2}
\end{figure}

Fig. \ref{fig:convergence2} presents the system cost by both the proposed scheme and the existing one with different numbers of MTCDs. Simulation results reveal that the system cost with different numbers of MTCDs or different optimization policies tends to be stable in each time frame. For instance, under the proposed scheme, when the number of MTCDs in the virtual network is $15$, the system cost by the proposed scheme remains nearly $20$ within each time frame. The results reveal that the proposed scheme has a stable optimization performance.

\section{Conclusions and Future Work}\label{sec:Conclusion}
In this paper, we proposed a novel scheme to jointly optimize energy consumption and computation in virtualized cellular networks with M2M communications. In the proposed framework, MTCDs will access the corresponding virtual network according to their functions or QoS requirements. In addition, the MEC is proposed as a promising technology for executing computing tasks.  Furthermore, we introduced the SDN paradigm to manage and integrate diverse protocols and standards, such as the proposed WNV and MEC in M2M communication networks. Simulation results demonstrated the effectiveness of the proposed framework. Future work is in progress to consider  delay and packet loss with M2M communications proposed in our framework.\

\balance
\bibliographystyle{IEEEtran}
\bibliography{limengreference}

\end{document}